\documentclass{bmvc2k}
\usepackage[misc]{ifsym}

\title{SALT: Parameter-Efficient Fine-Tuning via Singular Value Adaptation with Low-Rank Transformation}

\addauthor{Abdelrahman Elsayed* \Letter}{abdelrahman.elsayed@mbzuai.ac.ae}{1}
\addauthor{Sarim Hashmi*}{sarim.hashmi@mbzuai.ac.ae}{1}
\addauthor{Mohammed Elseiagy}{mohammed.abdelaziz@mbzuai.ac.ae}{1}
\addauthor{Hu Wang}{hu.wang@mbzuai.ac.ae}{1}
\addauthor{Mohammad Yaqub}{mohammad.yaqub@mbzuai.ac.ae}{1}
\addauthor{Ibrahim Almakky}{ibrahim.almakky@mbzuai.ac.ae}{1}

\addinstitution{
Mohamed Bin Zayed University of \\
Artificial Intelligence,\\
Abu Dhabi, UAE \\
}

\runninghead{ELSAYED ET AL.}{SALT}

\usepackage{booktabs}  % For professional table rules
\usepackage{siunitx}   % For decimal alignment
\usepackage{amsmath}
\usepackage{amssymb}
\usepackage{etex}
\usepackage{mathtools}
\usepackage[textsize=tiny]{todonotes}

\usepackage{array}      % For advanced column formatting
\usepackage{xcolor}     % If color is used
\usepackage{graphicx}      % For including graphics
\usepackage{caption}  
\usepackage{subfigure}  % For subfigure and subtable

\usepackage{multirow}      % For multi-row cells in tables
\usepackage{adjustbox}     % For adjusting table sizes
\usepackage{enumitem}      % For customized lists
\usepackage[T1]{fontenc}
\usepackage{xcolor,colortbl}
\usepackage{graphicx,verbatim}
\definecolor{tmi_blue}{cmyk}{100,0.37,0.0,0.15}
\usepackage{hyperref}
\hypersetup{
    colorlinks=true,
    linkcolor=tmi_blue,
    citecolor=tmi_blue,
    urlcolor=tmi_blue
}
\begin{document}

\maketitle
\let\thefootnote\relax\footnotetext{* Equal contribution}

\begin{abstract}
The complex nature of medical image segmentation calls for models that are specifically designed to capture detailed, domain-specific features. Large foundation models offer considerable flexibility, yet the cost of fine-tuning these models remains a significant barrier. Parameter-Efficient Fine-Tuning (PEFT) methods, such as Low-Rank Adaptation (LoRA), efficiently update model weights with low-rank matrices but may suffer from underfitting when the chosen rank is insufficient to capture domain-specific nuances.
Conversely, full-rank Singular Value Decomposition (SVD) based methods provide comprehensive updates by modifying all singular values, yet they often lack flexibility and exhibit variable performance across datasets.
We propose SALT (Singular Value Adaptation with Low-Rank Transformation), a method that selectively adapts the most influential singular values using trainable scale and shift parameters while complementing this with a low-rank update for the remaining subspace.
This hybrid approach harnesses the advantages of both LoRA and SVD, enabling effective adaptation without relying on increasing the size or depth of the model.
Evaluated on 5 challenging medical datasets, ranging from as few as 20 samples to 1000, SALT outperforms state-of-the-art PEFT (LoRA and SVD) by 2\% to 5\% in Dice with only 3.9\% trainable parameters, demonstrating robust adaptation even in low-resource settings. The code for this paper is available at : \url{https://github.com/BioMedIA-MBZUAI/SALT}.
% \keywords{Singular Value Decomposition, Parameter-Efficient Fine-Tuning, LoRA, Foundation Models, Domain Adaptation.}
\end{abstract}

%-------------------------------------------------------------------------
\section{Introduction}
Medical image segmentation plays a pivotal role in modern healthcare by enabling precise delineation of anatomical structures, pathological regions, and functional tissues across diverse imaging modalities such as CT, MRI, X-ray, ultrasound, and endoscopy \cite{surveyseg,unet}. While Deep Learning (DL) has revolutionized this field, with architectures like U-Net \cite{unet} and nnU-Net \cite{isensee2018nnunetselfadaptingframeworkunetbased} achieving state-of-the-art (SOTA) performance through sophisticated designs and extensive data augmentation, these models often rely on large parameter counts, limiting their adaptability to new datasets or modalities. For instance, nnU-Net’s success hinges on dynamic architecture adjustments and aggressive augmentation pipelines, while DeepLabv3+ \cite{deeplabv3} employs computationally intensive components like spatial pyramid pooling.
Recent advances in foundation models have introduced specialized tools for diverse computer vision tasks. CLIP \cite{clip} and ALIGN \cite{align}, trained on large-scale image-text datasets, excel at cross-modal understanding for zero-shot image classification and open-set retrieval. In contrast, the Segment Anything Model (SAM) \cite{sam,sam2}, trained on billions of natural images, exhibits robust zero-shot generalization capabilities and initially raised hopes for universal segmentation. However, SAM’s performance on medical data has proven suboptimal, with studies \cite{samandmedicalimaging,samandmedicalimaging12} demonstrating its inferiority to specialized models like U-Net++ \cite{unet++}, particularly in Dice scores across 12 medical datasets. Even with prompt engineering, SAM’s zero-shot accuracy remains below fully supervised methods \cite{SAMmedical}, underscoring the domain gap between natural and medical images.

Recent efforts to adapt SAM to medical imaging, such as MedSAM \cite{medsam}, SAM Adapter \cite{samadapter}, and S-SAM \cite{ssam}, have made strides by incorporating medical training data, low-rank adapters, or SVD-based tuning. Yet challenges persist: PEFT methods must balance minimal parameter overhead, computational feasibility, and preservation of pre-trained knowledge while addressing unique medical imaging challenges like speckle noise, low contrast, and modality-specific artifacts. Promising approaches that arose to adapt foundational models, especially language models, like LoRA \cite{lora} and AdaLoRA \cite{adalora} approximate weight updates via low-rank matrices but struggle to capture both dominant and nuanced features, while SVD-based methods like PiSSA \cite{pissa} focus on principal singular vectors to capture the most significant data patterns, but this rigid prioritization can limit their adaptability to dynamic or fine-grained data characteristics. To address these limitations, we propose \textbf{S}ingular Value \textbf{A}daptation with \textbf{L}ow-Rank \textbf{T}ransformation (SALT), a novel PEFT framework that synergizes the strengths of SVD and low-rank adaptation. SALT selectively scales critical singular values to preserve foundational knowledge while applying trainable low-rank transformations to residual components, enabling efficient adaptation to medical domains with minimal parameter overhead. Evaluated on five challenging medical datasets, with sample sizes ranging from 20 to 1000 and demanding a high level of detailed segmentation, SALT consistently outperforms LoRA and SVD, achieving a 2\% to 5\% improvement in Dice score while utilizing only 3.9\% trainable parameters. Our contributions are threefold: 

\begin{itemize}[nolistsep]
\item We introduce SALT, a hybrid PEFT approach that adapts dominant singular values via SVD while applying low-rank residuals, enabling efficient domain adaptation.

 \item We demonstrate SALT's utility through extensive experiments on diverse medical datasets from various modalities, showing improvements over SOTA PEFT methods.

 \item We present a comprehensive study exploring how singular value distributions and trainable parameters' allocation impact domain-specific fine-tuning for different challenging medical datasets.
 % \item Extending SALT to the new SAM2 models and outperforming LoRA and SVD based PEFTs when applied to SAM2.
 \end{itemize}
\section{Methodology}
\begin{figure*}[t]
    \centering
  \includegraphics[width=\textwidth,trim={30 35 30 20},clip]{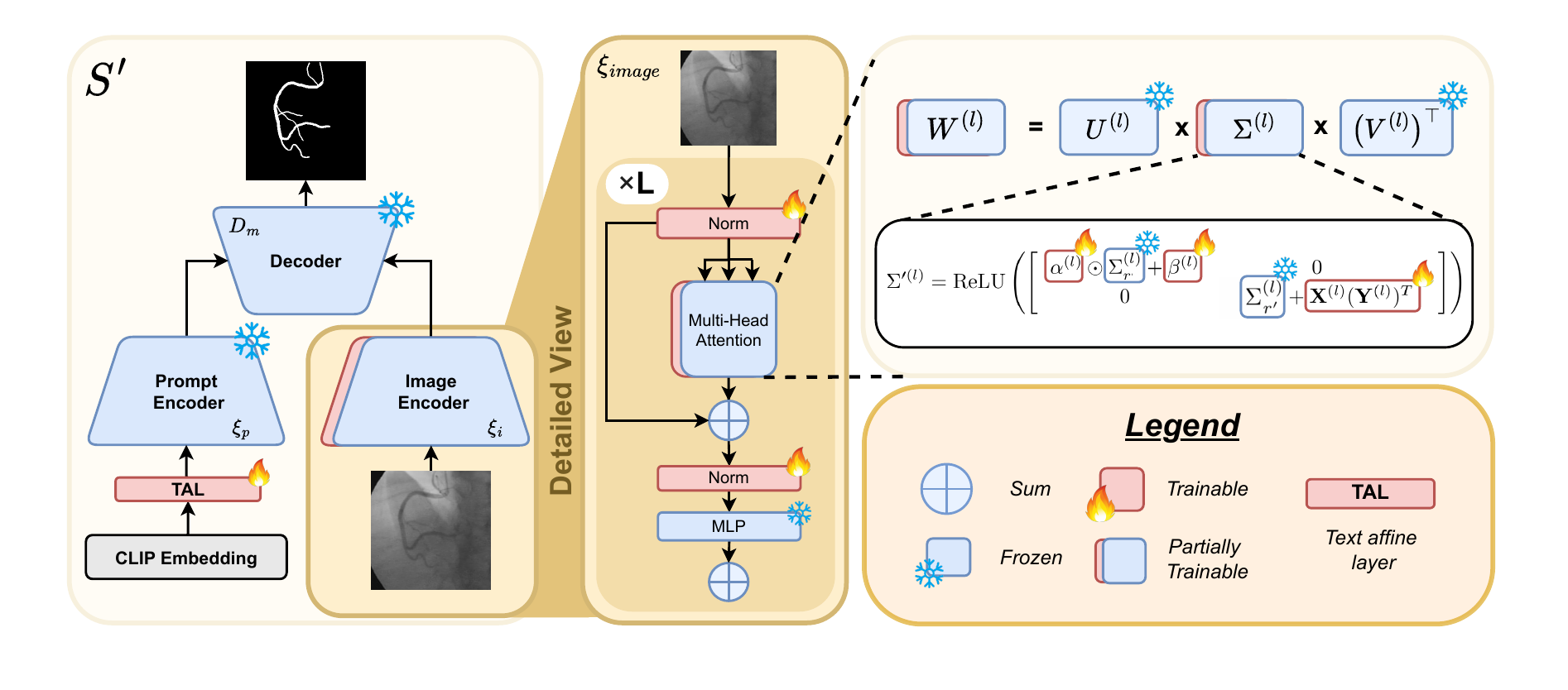}
    \caption{\textbf{Overview of the SALT architecture.} Starting from the segmentation model and providing a detailed view of the SALT mechanism within the Multi-Head Attention layers of the image encoder.}
    \label{fig:arch}
\end{figure*}
\subsection{Problem Formulation} Let S be a SAM model with pretrained parameters \( \Theta \), which consists of an image encoder (\(\mathcal{E}_{\mathrm{i}}\)), a prompt encoder (\(\mathcal{E}_{\mathrm{p}}\)), a mask decoder (\(\mathcal{D}_{\mathrm{m}}\)). As depicted in Fig.~\ref{fig:arch},  \(\mathcal{E}_{\mathrm{i}}\) contains \(L\) blocks, each comprising a Multi-Head Attention module (MHA) and a Multi-Layer Perceptron (MLP). For a given input image \(\mathbf{i}\), the encoder \(\mathcal{E}_{\mathrm{i}}\) produces intermediate feature maps \(\{\mathbf{f}^{(l)}\}_{l=1}^L\). Therefore, in the \(l\)-th block, the main operations can be written as:
\begin{equation}
q, k, v = W_{qkv}^l \, x^{(l-1)} + b_{qkv}^l, 
\quad
o = W_{M}^l \, x^{(l-1)} + b_{M}^l,
\end{equation}
where \(x^{(l-1)}\) is the input to block \(l\) (the output of block \(l\!-\!1\)), \(q,k,v\) are the query, key and value projection tensors produced by the MHA module in block \(l\) (that is, the inputs to the attention computation) and \(W_{qkv}^l, W_M^l \in \mathbb{R}^{D \times K}\) and \(b_{qkv}^l, b_M^l\in \mathbb{R}^{D}\) are the attention/MLP weights and biases for block \(l \in \{1,...,L\}\). While effective for natural images, the pre-trained weights \(W \in \mathbb{R}^{D \times K}\) may not capture all domain-specific nuances.
\subsection{Singular
Value Adaptation with Low-Rank Transformation}Through SALT, we refine these weights using the following procedure to obtain an adapted SAM model $S'$, shown in Fig.~\ref{fig:arch}. First,
consider a weight matrix \(\mathbf{W}^{(l)}\). SALT decomposes \(\mathbf{W}^{(l)}\) using Singular Value Decomposition (SVD).
\begin{equation}
\mathbf{W}^{(l)}= \mathbf{U}^{(l)}\,\boldsymbol{\Sigma}^{(l)}\,\bigl(\mathbf{V}^{(l)}\bigr)^\top,
\end{equation}
where \(\mathbf{U}^{(l)} \in \mathbb{R}^{D \times D}\) (left singular vectors), \((\mathbf{V}^{(l)})^\top \in \mathbb{R}^{K \times K}\) (right singular vectors), and \(\boldsymbol{\Sigma}^{(l)} \in \mathbb{R}^{D \times K}\) collects the singular values on its main diagonal. We denote $min(D,K)$ as $r_{max}$ and split this matrix into two matrices (i) $\boldsymbol{\Sigma}^{(l)}_r$ which contains a diagonal vector of top r singular values where \(r \leq r_{max}\) (ii)  $\boldsymbol{\Sigma}^{(l)}_{r'}$ containing the remaining $r'$ singular values where $r' = r_{max} - r$. Then SALT applies (i)  trainable scale and shift transformations to the top \(r\) singular values. Concretely, define parameters \(\alpha^{(l)}\) and \(\beta^{(l)}\) as  diagonal matrices \(\in \mathbb{R}^{r \times r}\) to transform $\boldsymbol{\Sigma}^{(l)}_r$:
\begin{equation}
\boldsymbol{\Sigma}^{'(l)}_r = \alpha^{(l)} \odot \boldsymbol{\Sigma}^{(l)}_r + \beta^{(l)},
\end{equation}
where \(\odot\) denotes Hadamard product, and (ii) a LoRA-based low-rank update \(\mathbf{X}^{(l)}\mathbf{Y}^{(l)\top}\) where $ \mathbf{X}^{(l)} \in  \mathbb{R}^{r' \times d_{\text{lora}}}$ , $ \mathbf{Y}^{(l)} \in  \mathbb{R}^{d_{\text{lora}}  \times r'}$ to the remaining singular values $\boldsymbol{\Sigma}^{(l)}_{r'}$. The updated core matrix \(\boldsymbol{\Sigma}^{'(l)}\) is then activated elementwise by \(\mathrm{ReLU}\) to ensure nonnegative entries, yielding the updated weight matrix:
\begin{figure}[t]
    \centering
  \includegraphics[width=\textwidth,trim={20 575 30 20},clip]{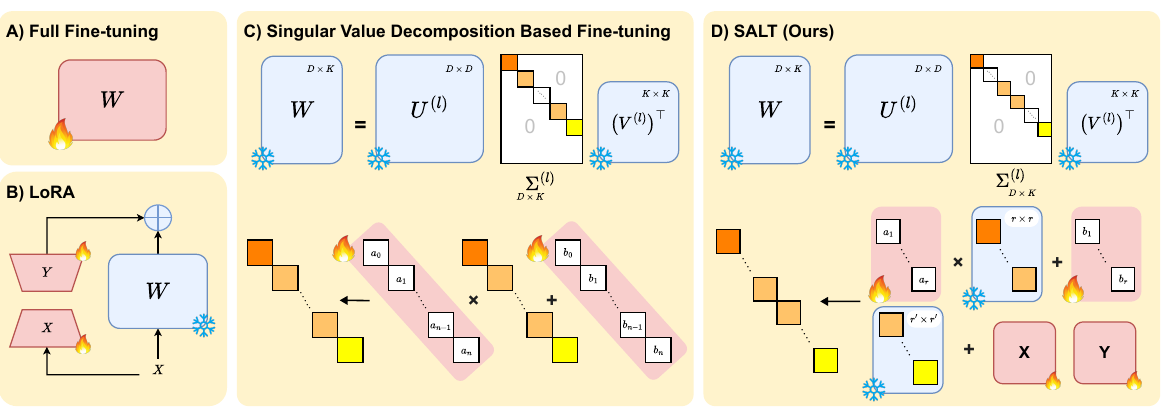}
       \caption{Comparison of PEFT methods: (A) Full fine-tuning updates all \(W\) parameters. (B) LoRA updates low-rank matrices. (C) SVD fine-tunes scale and shift components of \(W\). (D) SALT (Ours) integrates scale, shift, and low-rank updates for singular values.}
    \label{fig:comp}
\end{figure}
\begin{equation}
\widetilde{\mathbf{W}}^{(l)} \;=\; \mathbf{U}^{(l)}\,\bigl[\mathrm{ReLU}\!\bigl({\boldsymbol{\Sigma}}^{'(l)}\bigr)\bigr]\,\bigl(\mathbf{V}^{(l)}\bigr)^\top.
\end{equation}
During fine-tuning, only the newly introduced SALT parameters $\{\alpha^{(l)},\beta^{(l)},\mathbf{X}^{(l)}$, $\mathbf{Y}^{(l)}\}$ are optimized along with the normalization layers and the Text Affine Layer (TAL), while all original SAM weights remain frozen. Unlike the SVD’s $\Sigma^{(l)}$, $\Sigma'^{(l)}$ is not necessarily diagonal due to the added  \(\mathbf{X}^{(l)}\mathbf{Y}^{(l)\top}\) block. Embeddings of the text prompts are acquired using a frozen CLIP model \cite{clip}, after which TAL projects the CLIP embeddings to the mask decoder (\(\mathcal{D}_{\mathrm{m}}\)) token space \cite{ssam}. The prompt encoder \(\mathcal{E}_{\mathrm{p}}\), processes the projected CLIP embeddings \(p\), and merges them with the image-encoder \(\mathcal{E}_{\mathrm{i}}\) output in the mask decoder \(\mathcal{D}_{\mathrm{m}}\). We define SALT, denoted \(S'\), as the combination of two components: The original SAM model's frozen output \(S(i; \Theta_{\text{fixed}})\), which remains static during fine-tuning, and a lightweight, learnable adjustment term \(\Delta S(i; \theta)\), introduced by SALT layers, capturing task-specific refinements through the trainable parameters \(\theta\). This formulation ensures that the core SAM backbone remains unchanged, while task-specific adaptations are layered on top via the SALT modules. Putting this together, our objective is to minimize the loss function $ \mathcal{L}(S'(i),y)$
where \( i \) is an input sample and \( y \) is the corresponding ground truth label.

\noindent SALT performs the SVD factorization only once per layer at initialization; after that, all training steps update just the low-dimensional scale, shift, and LoRA parameters, keeping the computational overhead low. To train SALT, we use a combination of segmentation losses, including a Focal loss, a Dice loss, and a regularization term expressed as: 
\begin{equation}
\mathcal{L}_{\mathrm{reg}}
= \sum_{l=1}^{L}
\Big\|
\alpha^{(l)} \odot  \boldsymbol{\Sigma}^{(l)}_{r}
+ \beta^{(l)}
- \boldsymbol{\Sigma}^{(l)}_{r}
\Big\|_{F}
\;+\;
\big\| \mathbf{X}^{(l)} \mathbf{Y}^{(l)} \big\|_{F}
\end{equation}
where \(\|\cdot\|_F\) denotes the Frobenius norm. This dual constraint balances two goals: retaining the useful, pre-trained features and making only the minimal modifications needed for effective domain adaptation.\\

\textbf{Rationale for the SALT Design}
As shown in Fig. \ref{fig:comp}, LoRA \cite{lora} modifies a weight matrix \(W \in \mathbb{R}^{D \times K}\) by adding \(XY\), where \(X \in \mathbb{R}^{D \times r}\) and \(Y \in \mathbb{R}^{r \times K}\) have rank \(r \ll \min(D, K)\). This reduces trainable parameters from \(D \times K\), down to \(r(D + K)\), but can underfit if \(r\) is too small. By contrast, SVD-based fine tuning in S-SAM \cite{ssam},  shown in Fig. \ref{fig:comp}, learns only the \emph{scale} and \emph{shift} of each singular value in \(W\). Since \(\mathrm{rank}(W)=\min(D,K)\), this corresponds to \(2\times \min(D,K)\) parameters, fewer than LoRA, yet fully updating every singular value may lack flexibility for subtle domain shifts.
\textbf{SALT} in Fig. \ref{fig:comp} merges the strengths of both. It trains the top singular values via scale and shift and uses LoRA blocks for the lower singular subspace. This dual strategy retains key features while capturing domain-specific nuances, balancing parameter efficiency with adaptation capability.
\section{Experiments}
\subsection{Datasets}
In this work, we use datasets spanning three anatomical domains: neurovascular analysis with \emph{DIAS} \cite{dias}, a dynamic 2D+time DSA dataset for intracranial artery segmentation in cerebrovascular disease (20 train/10 test). Retinal assessment via \emph{ROSE} \cite{rose} for OCT-Angiography-based microvasculature analysis (22 train/8 test). \emph{DRIVE} \cite{drive} for diabetic retinopathy screening in RGB fundus images (14 train/6 test). Cardiovascular studies using \emph{ARCADE} \cite{arcade} (700 train/300 test) and \emph{XRay-Angio} \cite{xrayangio} (93 train/41 test), both leveraging X-ray angiography, where the former is for region-based coronary artery disease diagnostics and the latter is for multiscale segmentation of occluded vessels. The chosen datasets span three organ systems (brain, eye, heart), four imaging modalities (DSA, OCT-A, RGB fundus, X-ray), and both temporal and spatial resolutions. This ensures that our evaluation is robust across diverse segmentation challenges. Moreover, the diversity in dataset sizes offers a robust testbed for evaluating how different PEFT methods cope with varying dataset scales, ranging from low-resource to large-scale settings, assessing their adaptability under challenging conditions.
% Highlight the varying number of samples between the datasets and the importance of this to the effective testing of PEFT methods. 
The datasets were randomly partitioned into training and testing subsets to ensure unbiased evaluation.
\subsection{Implementation Details.} 
We apply SALT to the image encoder layers of the prompt-adapted SAM with a 512 input size and text-based prompts. Following the pipeline in S-SAM \cite{ssam}, training uses AdamW (\(\text{LR} = 1 \times 10^{-4}\), weight decay \(= 1 \times 10^{-2}\)), batch size 5, and 200 epochs, with a step decay scheduler for learning rate adjustment. For simplicity and consistency across datasets, we used a single text prompt \emph{``Vein''}, reflecting the primary anatomical structure of interest across all datasets and avoiding dataset-specific prompt engineering. Data augmentation includes random horizontal flips (50\% probability), rotations up to $\pm 10^\circ$ (50\% probability), saturation adjustments (20\% probability), and brightness scaling with random intensity (50\% probability), enhancing robustness to geometric and photometric variations. Images and masks are resized with anti-aliasing and padded to maintain consistent input size, while random cropping ensures the presence of foreground pixels during training. Input images are normalized using SAM’s expected mean and standard deviation to align with the pretrained encoder distribution. The total loss combines focal loss, Dice loss, and regularization loss (scaled by 0.01). Evaluation metrics include the Dice Similarity Coefficient (DSC) and 95th percentile Hausdorff Distance (HD95). Experiments were conducted on an NVIDIA RTX 4090 (24GB). We follow the same experimental setup to train our traditional baseline models, including a standard U-Net \cite{unet} with a 5-block encoder-decoder architecture initialized with 32 feature maps, SegNet \cite{segnet} with its original architecture, Segformer \cite{segformer}, and the other baseline methods.

\section{Results and Discussion}
\subsection{Main Results}
\begin{table}[t]
    \centering
    \caption{Performance comparison of traditional DL and SAM-based PEFT methods using Dice and HD95 metrics on a variety of medical segmentation datasets.\\}  
    \label{tab:performance_comparison_hd95_std}

    \resizebox{\textwidth}{!}{%
    \begin{tabular}{l c c | cc | cc | cc | cc | cc | cc}
        \toprule
        \textbf{Model} & \textbf{Rank} & \textbf{\%Trainable} 
        & \multicolumn{2}{c|}{\textbf{ARCADE}} 
        & \multicolumn{2}{c|}{\textbf{XRay-Angio}} 
        & \multicolumn{2}{c|}{\textbf{DIAS}} 
        & \multicolumn{2}{c|}{\textbf{ROSE}} 
        & \multicolumn{2}{c|}{\textbf{DRIVE}} 
        & \multicolumn{2}{c}{\textbf{Average}} \\

        \cmidrule(lr){4-5}
        \cmidrule(lr){6-7}
        \cmidrule(lr){8-9}
        \cmidrule(lr){10-11}
        \cmidrule(lr){12-13}
        \cmidrule(lr){14-15}

           &   & 
           & \textbf{Dice $\uparrow$} & \textbf{HD95 $\downarrow$} 
         & \textbf{Dice $\uparrow$} & \textbf{HD95 $\downarrow$} 
         & \textbf{Dice $\uparrow$} & \textbf{HD95 $\downarrow$}  
         & \textbf{Dice $\uparrow$} & \textbf{HD95 $\downarrow$} 
         & \textbf{Dice $\uparrow$} & \textbf{HD95 $\downarrow$}  
         & \textbf{Dice $\uparrow$} & \textbf{HD95 $\downarrow$} \\
        \midrule
        \multicolumn{15}{c}{\textbf{Traditional DL Methods}} \\
        \midrule
        U-Net \cite{unet} 
        & -- & 100\% / 6.5M
        & 0.36 & 157.14   % ARCADE
        & 0.71 & 30.40    % XRay-Angio
        & 0.13 & 262.66   % DIAS
        & 0.63 &  8.07    % ROSE
        & 0.18 & 84.43    % Drive
        & 0.40 & 108.54   % Avg
        \\

        UNETR \cite{unetr} 
        & -- & 100\% / 91M
        & 0.48 & 162.44 
        & 0.74 &  20.62
        & 0.31 & 183.97 
        & 0.63 &   8.96
        & 0.46 &  33.58
        & 0.52 &  81.91
        \\

        RegUNet \cite{unet}
        & -- & 100\% / 7.8M
        & 0.11 & 247.79
        & 0.42 &  79.51
        & 0.35 & 172.09
        & 0.22 &  58.08
        & 0.43 &  72.97
        & 0.31 & 126.08
        \\

        SegNet \cite{segnet}
        & -- & 100\% / 29M
        & 0.14 & 231.11
        & 0.09 & 110.94
        & 0.13 & 295.06
        & 0.38 &   9.09
        & 0.14 &  79.03
        & 0.18 & 145.05
        \\

        DeepLabV3+ \cite{deeplabv3}
        & -- & 100\% / 45M
        & 0.55 & 103.22
        & 0.74 &  \textbf{14.03}
        & 0.55 &  77.66
        & 0.52 &  12.90
        & 0.46 &  27.79
        & 0.56 &  47.12
        \\

        Segformer \cite{segformer}
        & -- & 100\% / 43M
        & 0.53 & 113.98
        & 0.75 &  16.55
        & 0.57 &  73.64
        & 0.56 &  11.21
        & 0.48 &  36.98
        & 0.58 &  50.47
        \\
        \midrule
        \multicolumn{15}{c}{\textbf{SAM-Based PEFT Methods} (241 M)} \\
        \midrule
        SAM w/ prompts & -- & -- 
        & 0.00 &  – 
        & 0.04 &  – 
        & 0.00 &  – 
        & 0.00 &  – 
        & 0.00 &  – 
        & 0.01 &  – 
        \\

        MedSAM & -- & -- 
        & 0.18 & 115.18
        & 0.13 &  80.05
        & 0.29 & 168.82
        & 0.32 & 131.89
        & 0.21 & 153.23
        & 0.23 & 129.83
        \\

        S-SAM \cite{ssam}
        & --  & 0.40\% / 1M
        & 0.76 & 52.44
        & 0.75 & 27.56
        & 0.67 & 42.36
        & 0.65 & 12.41
        & 0.70 & 15.81
        & 0.71 & 30.12
        \\

        LoRA \cite{lora}
        & 4   & 0.64\% / 1.5M
        & 0.78 & 39.58
        & 0.73 & 26.78
        & 0.61 & 27.81
        & 0.59 & 16.26
        & 0.62 & 21.04
        & 0.66 & 26.29
        \\

        LoRA \cite{lora}
        & 256 & 14.08\% / 33.9M
        & \textbf{0.81} & \textbf{36.89}
        & 0.76 & 24.84
        & 0.63 & 34.36
        & 0.62 & 14.62
        & 0.68 & 19.00
        & 0.70 & 25.94
        \\

        \rowcolor{tmi_blue!10}
        SALT (ours)
        & 4   & 0.46\% / 1.1M
        & 0.78 & 51.74
        & 0.75 & 29.55
        & 0.69 & 39.58
        & 0.66 & 12.19
        & 0.72 & 16.04
        & 0.72 & 29.82
        \\

        \rowcolor{tmi_blue!10}
        SALT (ours)
        & 256 & 3.9\% / 9.4M
        & \textbf{0.81} & 42.50
        & \textbf{0.77} & 23.67
        & \textbf{0.71} & \textbf{27.35}
        & \textbf{0.67} & 12.64
        & \textbf{0.75} & \textbf{13.20}
        & \textbf{0.74} & \textbf{23.87}
        \\
        \bottomrule
    \end{tabular}%
    }
\end{table}
We evaluate the performance of SALT against SOTA PEFT techniques, specifically LoRA \cite{lora} and S-SAM \cite{ssam}. As shown in Table \ref{tab:performance_comparison_hd95_std}, our experimental evaluation demonstrates the effectiveness of SAM-based PEFT methods compared to traditional deep learning approaches across five medical imaging datasets. While conventional methods like DeepLabV3+ \cite{deeplabv3} and Vanilla U-Net, \cite{unet} showed competitive performance on datasets such as XRay-Angio \cite{xrayangio}, they require training of all parameters. Additionally, we evaluated MedSAM\cite{medsam}, a universal medical segmentation foundation model trained on over 1.57 million image–mask pairs, under zero-shot conditions. It achieves only an average Dice score of 0.23 with an HD95 of 129.8, demonstrating that even a highly specialized backbone benefits greatly from in-domain fine-tuning. On the other hand, SALT achieves superior performance with significantly fewer trainable parameters. With rank-256 configuration, attaining an average Dice score of 0.74 while requiring only 3.9\% of parameters to be trained. This represents a substantial improvement over both traditional methods and other parameter-efficient approaches like LoRA \cite{lora} and S-SAM \cite{ssam}. Notably, SALT consistently achieves higher Dice scores over other methods across all datasets, achieving the highest scores on DIAS (0.71), ROSE (0.67), XRay-Angio (0.77), and DRIVE (0.75).
The rank configuration comparison reveals that while rank-4 SALT delivers competitive results with minimal parameter overhead (0.46\% trainable), the rank-256 version provides optimal performance with a modest increase in trainable parameters.

Table \ref{tab:performance_comparison_hd95_std} quantifies the parameter and computational efficiency by comparing the number of trainable parameters for each method, providing a direct measure of tuning overhead and runtime cost. PEFT methods achieve state-of-the-art performance while significantly reducing the number of trainable parameters and computational demands. For instance, LoRA at rank 256 tunes 33.9M parameters (14\%), whereas SALT at the same rank requires only 9.4M (3.9\%). Both SALT and LoRA reduce overhead further to approximately 0.5\% at rank 4. Furthermore, LoRA's GFLOPs\footnote{\label{fn:compute}\textbf{GFLOPs.} We measure per-sample MACs (eval, no-grad) and report GFLOPs using \(1\,\mathrm{MAC}=2\,\mathrm{FLOPs}\) (\(\mathrm{GFLOPs}=2\,\mathrm{MACs}/10^{9}\)). Adapter overheads are computed by our custom hooks.} grow roughly linearly with rank. At rank 4, it adds 1.24 GFLOPs while at rank 256 it adds 79.2 GFLOPs, a x64 jump matching the rank increase. By contrast, SALT is nearly rank-independent because the weight recomposition step dominates. With (0.46\%) trainable parameters, SALT adds 22.88 GFLOPs. At a higher rank of 256 and (3.9\%) trainable parameters, this increases only by 0.64 GFLOPs to reach 23.52 GFLOPs. We exclude the one-off SVD factorization from per-sample GFLOPs, since it is incurred only once. For completeness, the total one-time setup across all SVD/SALT layers is 871.8 GFLOPs, which amortizes to $\approx 4.36$ GFLOPs per epoch over 200 epochs. 

Performance gains also correlate with dataset size. LoRA tends to perform well on larger datasets like ARCADE \cite{arcade} and XRay-Angio\cite{xrayangio} but experiences degradation on smaller datasets. In contrast, SALT consistently delivers strong and stable performance across all dataset sizes, demonstrating high robustness, especially in low-resource scenarios. This can be attributed to SALT’s targeted adaptation, where scale and shift-guided updates to the top singular values preserve useful pre-trained features, serving as a built-in regularizer that prevents overfitting on small datasets while residual low-rank updates simultaneously enhance model capacity.

Analysis of boundary accuracy via HD95 further underscores SALT's performance. While traditional methods exhibited high HD95 errors, such as SegNet \cite{segnet} scoring 145.05 and RegUnet \cite{unet} scoring 126.08, SAM-based PEFT approaches showed improvements, with SALT achieving the lowest average HD95 ($23.87$) against LoRA\cite{lora} ($25.94$) and S-SAM \cite{ssam} ($30.12$). Notably, SALT demonstrated strong boundary adherence in challenging cases. SALT's 13.20 HD95 on DRIVE \cite{drive} against S-SAM \cite{ssam} achieving 15.81 reflects precise vessel delineation in low-contrast retinal images. This dual advantage in both Dice and HD95 confirms SALT's ability to maintain structural accuracy and boundary precision while operating in parameter-efficient regimes.

Explaining the gain in performance, and drawing from the Eckart–Young–Mirsky theorem \cite{Eckart1936}, SALT can be interpreted as adjusting the most informative components of the weight matrices while bounding operator-norm deviations. This would help maintain stability during adaptation. From a domain adaptation perspective, these updates would reduce the covariate shift between source (natural image) and target (medical) distributions by aligning feature covariances in the dominant subspace, while the residual low-rank updates would contribute to conditional alignment.

A closer inspection reveals that boundary improvements do not always scale proportionally with Dice despite consistent gains (Table \ref{tab:performance_comparison_hd95_std}). We observe cases where boundary precision (HD95) does not improve at the same rate. 
For example, on ARCADE, SALT with rank 256 matches LoRA's Dice score (0.81) but has a higher HD95 (42.50 vs.\ 36.89). At the lower rank of 4, SALT improves Dice on DIAS (0.69 vs.\ 0.61) yet underperforms on HD95 (39.58 vs.\ 27.81).
These observations might suggest that SALT’s dominant-singular-value rescaling and low-rank residuals may prioritize region overlap while occasionally under-capturing high-frequency boundary cues, particularly at lower ranks.

\subsection{Qualitative Analysis}
\begin{figure*}[t!]
    \centering \includegraphics[width=0.95\textwidth,trim={25 350 5 30},clip]{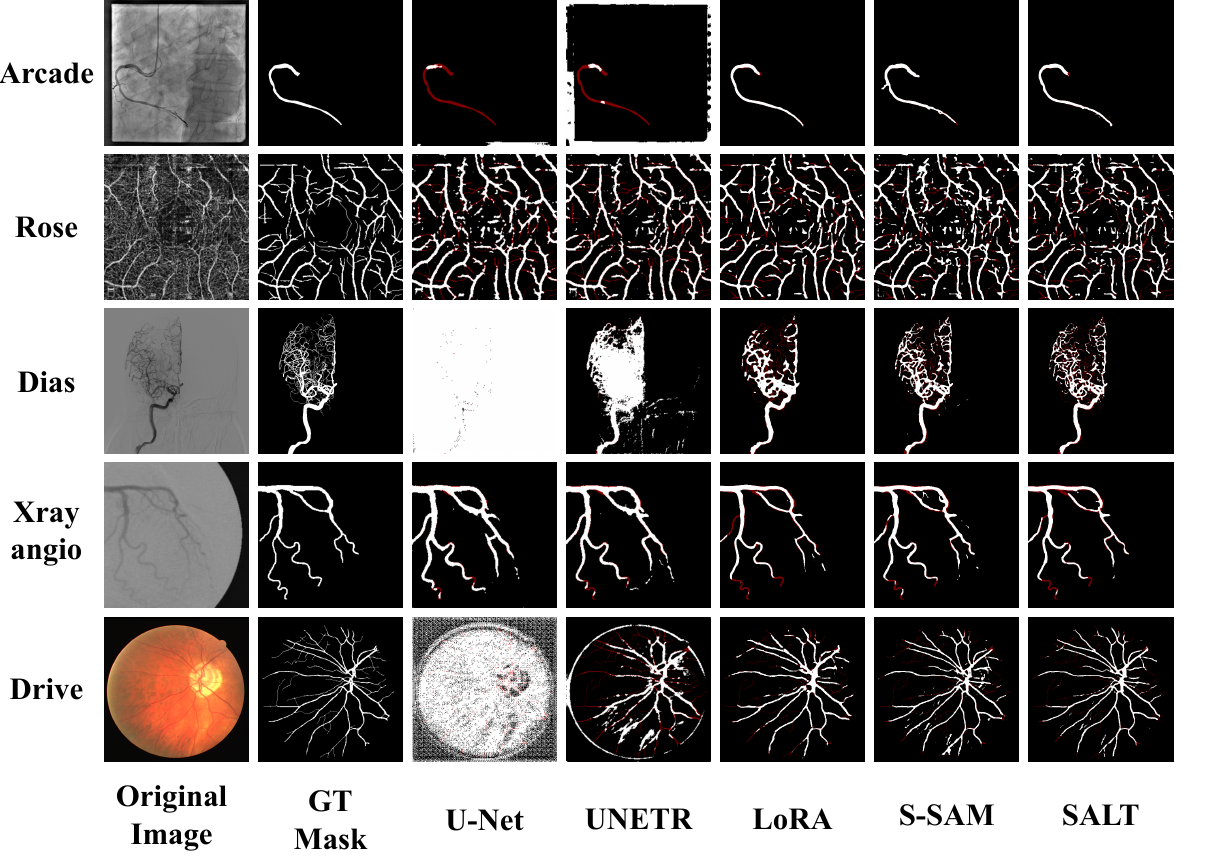}
       
\caption{%
\textbf{Qualitative Analysis.} The figure displays segmentation results for five datasets (Arcade \cite{arcade}, ROSE \cite{rose}, DIAS \cite{dias}, XRay-Angio \cite{xrayangio}, DRIVE \cite{drive}). Columns show the ground truth, SALT,  SAM-based PEFT methods, and traditional DL models. Predicted masks are white, and false negatives are red. %\hu{The names on the left side are not complete}}
       }
    \label{fig:qualitiative}
\end{figure*}
As illustrated in Fig. \ref{fig:qualitiative}, the qualitative results highlight the SOTA performance of our proposed SALT method across various medical imaging datasets. Compared to traditional approaches like UNETR \cite{unetr} and U-Net \cite{unet}, as well as other SAM-based PEFT methods such as S-SAM \cite{ssam} and LoRA \cite{lora}, SALT consistently delivers more precise vessel segmentation. In the ROSE \cite{rose} dataset, SALT excels at capturing the intricate, web-like patterns of retinal vessels while maintaining structural continuity with fewer false positives. For the DIAS \cite{dias} dataset, SALT effectively segments complex branching patterns, preserving both main vessel trunks and subtle secondary branches. In X-ray angiography \cite{xrayangio} images, SALT achieves clear vessel boundaries with fewer false positives compared to traditional methods. Additionally, in the DRIVE \cite{drive} dataset, it successfully segments retinal vessels despite varying widths and complex backgrounds.\\
% \subsection{Extension to SAM2 \cite{sam2}.} 

\subsection{SALT with SAM2} 
\begin{figure*}[t]
    \centering
    \subfigure[]{
        \includegraphics[width=0.45\textwidth,trim={0 0 0 0},clip]{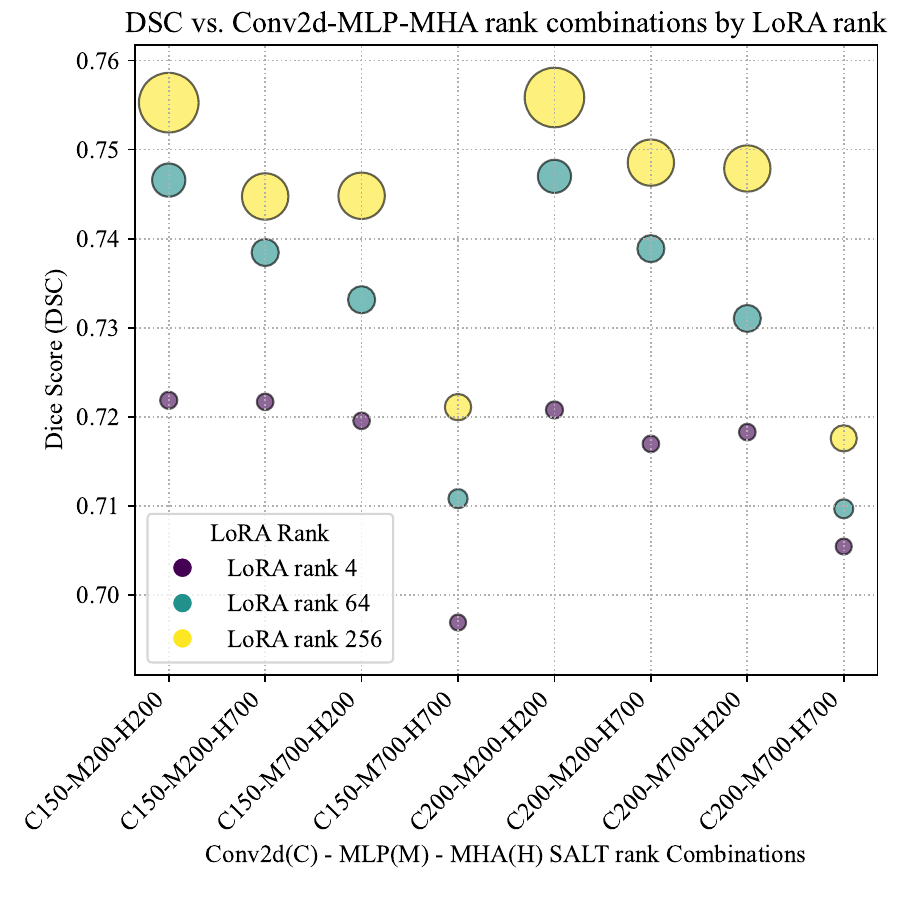}
        \label{fig:ablation}
    }
    \hfill
    \subfigure[]{
        \includegraphics[width=0.50\textwidth,trim={135 220 120 190},clip]{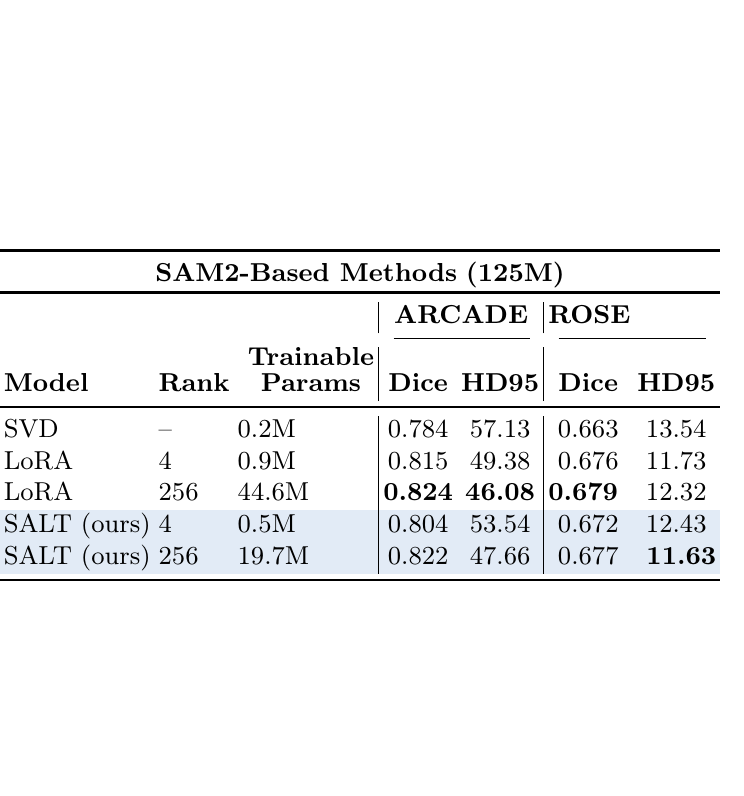}
        \label{fig:comparison}
    }
    \caption{%
    (a) Model performance and parameter efficiency across different LoRA and SALT ranks: x-axis shows parameter combinations for different layers, y-axis shows the Dice score, and the circle size indicates the percentage of trainable parameters.  
    (b) Comparison of SALT, LoRA \cite{lora}, and SVD-based PEFT on SAM2 over two datasets of varying sizes.
    }
    \label{fig:combined_analysis}
\end{figure*}
We further evaluate SALT on SAM2, comparing with LoRA \cite{lora}, and SVD-based fine tuning on the Hiera \cite{hiera} image encoder. The SAM2 \cite{sam2} image encoder utilizes a hierarchical architecture, capturing multi-scale features from input images with variable embedding dimensions ranging from low-dimensional, early-stage representations to high-dimensional, later-stage ones. This design enables LoRA to adapt effectively. In contrast, the original SAM \cite{sam} image encoder employs a backbone with constant, high-dimensional embeddings, where LoRA exhibited instability and was sometimes outperformed by S-SAM's SVD-based fine-tuning method \cite{ssam}. In SAM2, however, LoRA consistently outperforms relative to SVD, while SALT achieves competitive performance to LoRA with 2.4 times fewer parameters at rank 256 and 1.8× fewer at rank 4. According to the table shown in figure \ref{fig:comparison}, SALT, at a LoRA rank configuration of 4, achieves a Dice score of 0.804 on the ARCADE \cite{arcade} dataset with only 0.5 million trainable parameters, substantially reducing the parameter footprint compared to LoRA’s 0.9 million parameters. When scaled to a higher rank of 256, SALT attains Dice scores of 0.822 and 0.677 on the ARCADE \cite{arcade} and ROSE \cite{rose} datasets respectively, closely matching the performance of LoRA (0.824 and 0.679) despite requiring less than half the number of trainable parameters (19.7 million versus 44.6 million). for boundary analysis, SALT achieves superior boundary segmentation on the ROSE dataset as indicated by the lowest HD95 value of 11.63, outperforming both LoRA and the baseline SVD fine-tuning. Collectively, these results show that SALT offers an effective balance between parameter efficiency and segmentation accuracy for the SAM2 architecture, enabling precise and resource-efficient fine-tuning through optimized parameter allocation.
% \subsection{Computational Efficiency}
% \subsection{Ablation Study.}
\subsection{Ablation Study}
We conducted comprehensive ablation experiments on the DRIVE dataset \cite{drive} to evaluate the impact of different architectural components and rank configurations in SALT. Fig. \ref{fig:ablation} presents the results across various combinations of LoRA ranks (4, 64, 256) and SALT ranks for MHA, neck convolutional layers (NECK-Conv2d), and MLP modules. The SALT rank for a given layer refers to the number of dominant singular values in the layer’s weight matrix that are adaptively scaled and shifted during fine-tuning. For example, a SALT rank of 200 for the  Multi-Layer Perceptron modules means that for each weight matrix, we update the top 200 singular values using scale and shift parameters, while the rest are updated using the low-rank updates. Key findings are: 1) increasing the LoRA rank from 4 to 256 consistently improved Dice scores ($0.72\ \rightarrow  0.76 $) while maintaining a modest parameter increase from 0.46\% to 5.86\% and 2) reducing MHA and MLP ranks from 700 to 200 further enhanced performance, increasing Dice scores from 0.75 to 0.76 with LoRA rank 256. These results suggest that higher LoRA ranks improve expressiveness, while component-specific SALT rank selection is crucial for optimal performance. By decreasing the SALT rank, we allow less dominant singular values to be adapted through LoRA updates, which enhances the model's adaptation capacity. The optimal configuration (LoRA rank 256 and components rank 200) achieved a Dice score of 0.76 with only 5.84\% of trainable parameters, demonstrating SALT's effective balance of performance and efficiency.
\section{Conclusion}
In this paper, we introduce SALT, a novel PEFT framework that adapts the Segment Anything Model (SAM) to domain-specific tasks, particularly medical image segmentation. By utilizing SVD-based scaling and shifting of the top singular values with LoRA-based low-rank updates, SALT achieves efficient and flexible adaptation, enabling precise segmentation with minimal computational overhead. Experiments show that SALT outperforms state-of-the-art SAM PEFT adaptations while preserving parameter efficiency and adaptability.
While we have focused on SAM and its variants, many other large‐scale vision backbones, Vision Transformers (ViTs), and even multi‐modal encoders could benefit from SALT. In future work, applying SALT to these architectures will test its versatility across different embedding dimensions, attention patterns, and domain gaps. Moreover, the current design fixes the selection of ranks and SALT hyperparameters per layer. In future work, we aim to develop methods to automatically choose each layer’s rank to tailor parameter allocation to each layer’s intrinsic complexity. Finally, we will add boundary-aware training and small multi-scale tweaks to better capture fine edges and lower HD95, especially when data is scarce or ranks are very low.

\bibliography{egbib}
\end{document}